\documentclass{article}
\usepackage{amssymb}

\newcommand{\beq}{\begin{equation}}
\newcommand{\eeq}{\end{equation}}
\newcommand{\beqn}{\begin{eqnarray}}
\newcommand{\eeqn}{\end{eqnarray}}

\newtheorem{defn}{Definition}
\newtheorem{cor}{Corollary}
\newtheorem{exmp}{Example}
\newtheorem{thm}{Theorem}
\def\tr{{\rm tr}\,}
\def\dST{\displaystyle}

\title{\bf Relations Between  Quantum Maps and Quantum States}
\author{M. Asorey \\[1ex]
Departamento de F\'{\i}sica Te\'orica,  Universidad de Zaragoza\\
50009 Zaragoza, Spain\\
e-mail: asorey@saturno.unizar.es\\[1ex]
A.  Kossakowski \\[1ex]
Institute of Physics, Nicolaus Copernicus University \\
Toru\'n 87\,--\,100, Poland \\
e-mail: kossak@phys.uni.torun.pl\\[1ex]
G. Marmo\\[1ex]
Dipartimento  di Scienze Fisiche, Universit\'a
Federico II di Napoli \\ and \\ INFN, Sezione di Napoli\\
Complesso Univ.\ di Monte Sant'Angelo, Via Cintia, 80125 Napoli, Italy \\
e-mail: marmo@na.infn.it\\[1ex]
E.\,C.\,G. Sudarshan\\[1ex]
Department of Physics, University of Texas at Austin \\
Austin, Texas 78712-1081 \\
e-mail: sudarshan@physics.utexas.edu}

\begin{document}
\maketitle

\begin{abstract}
The relation between completely positive maps and compound
states is investigated in terms of the notion of quantum conditional
probability.
\end{abstract}

\section{Introduction}

In quantum information two problems play a relevant role, the
first one concerns the study of the dynamical change of states of a system
by means of completely positive maps,  commonly
called channels, the second one is to describe correlations
between the initial and final states; such correlations are
described by compound states.

The connection between the two concepts is based on a very general
principle. Indeed, in any Hilbert space ${\mathcal H}$ there is an one-to-one correspondence
between the set ${\mathcal P}^p_q$ of $p$-contravariant $q$-covariant tensors and
the set  ${\mathcal P}_{p+q}$ of $p+q$ covariant tensors. The equivalence is due to
the identification of  ${\mathcal H}$ and its dual space ${\mathcal H}^\ast$ by means of the
hermitian product. Consequently,  endomorphisms of ${\mathcal H}$ which are in  ${\mathcal P}^1_1$ are in
one-to-one correspondence with  $2$-covariant tensors in ${\mathcal P}_2$.  In
particular, if we consider the Hilbert spaces of  Hilbert-Schmidt operators on
${\mathcal H_1}$ and ${\mathcal H_2}$,  any map of  the Hilbert-Schmidt operators on  ${\mathcal H_1}$
into those on ${\mathcal H_2}$ is associated to a state on the tensor product of
the spaces of Hilbert-Schmidt  operators on ${\mathcal H_1}$ and ${\mathcal H_2}$.

The  correspondence can also be formulated at $C^*$-algebraic
level. However, only the finite-dimensional case will be
considered here.

Let us consider two systems described by $(M_n,\, S(M_n))$ and
$(M_m,\, S (M_m))$. The first one describes an initial ({\it input})
system and the second one a final ({\it output}) system. The symbol
$M_n$ stands for the algebra of $n \times n$ complex matrices and
the symbol $S(M_n)$ stands for the set of all states on $M_n$, i.e.\
the set of all density matrices. Moreover, $I_n$ denotes the
identity matrix in $M_n$. Let us consider a map $\varphi^* \colon  S
(M_n) \to S (M_m)$, such that its dual map $\varphi  \,\colon  M_m
\to M_n$ is completely positive and normalized, i.e.\ $\varphi (I_m)
= I_n$.

For an initial state $\rho \in M_n$ and final state $\varphi^*
(\rho) \in M_m$, a composite state $\omega \in S (M_n \otimes
M_m)$ should satisfy the following two conditions:
\begin{itemize}
\item[i)] \quad $ \omega (a \otimes I_m) = \rho (a)$, for all $a \in M_n$
\item[ii)] \quad $ \omega (I_n  \otimes b) = \varphi^\ast(\rho) (b)$ for
all $b \in M_m$.
\end{itemize}

It is well known that joint probability measures do not
generally exist for quantum systems, therefore it is difficult to
define a compound state $\omega$ satisfying the above conditions.

The first construction of a compound state $\omega$ satisfying the
above two conditions has been given by Ohya [1,\,2]. Let $\rho \in
M_n$, then $\rho$ has the following spectral decomposition
\beq
\rho \;=\; \sum_k \lambda_k m_k \rho_k\,,
\eeq
where
\beq \rho_k \;=\;
\frac{1}{m_k} p_k\, , \qquad m_k \;=\; \mbox{Tr}\, p_k\, ,
\eeq
$\lambda_k$
are the eigenvalues of $\rho$, and $p_k$ are eigenprojectors of
$\rho$, respectively.

Then for any $\varphi^* \colon S (M_n) \to S (M_m)$ the compound
state $\omega_\varphi \in S (M_n \otimes M_m)$ has the form \beq
\omega_\varphi \;=\; \sum_k  \lambda_k m_k\,  \rho_k \otimes
\varphi^* (\rho_k)\,. \eeq Let us observe that $\omega_\varphi$ is a
separable state on $M_n \otimes M_m$, $\varphi$ is a positive
normalized map $\varphi \,\colon  M_m \to M_n$, and the construction
of $\omega_\varphi$ is non-linear with respect to $\rho$.

We notice that the Ohya compound state can be constructed in the
case of general $C^*$-algebraic setting.

The construction of compound states which will be studied in the
present paper can be  described as follows.

Let $\sigma \in S (M_n \otimes M_m)$, and suppose that \beq
\mbox{tr}_{\mathbb{C}^m}\, \sigma \;=\; \sigma_1 \;>\;0\,. \eeq
Then, one can define the following operator
 \beqn
 &  \pi(\sigma) \, \colon \, \mathbb{C}^n
\otimes \mathbb{C}^m \;\longrightarrow\; \mathbb{C}^n \otimes
\mathbb{C}^m & \nonumber \\
&  \pi(\sigma) \;=\; (\sigma_1^{-1/2} \otimes I_m)\, \sigma\, (\sigma_1^{-1/2}
\otimes I_m)\,,& \eeqn
which has the properties
\beqn
& \pi(\sigma) \;\geq\; 0 &\\
& \mbox{tr}_{\mathbb{C}^m}\, \pi(\sigma) \;=\; I_m \,. & \eeqn
It follows from (1.4) and (1.5) that the operator $\pi$ is the quantum analogue of
classical conditional probability.
Another definition of quantum conditional probability has been given in \cite{cerf}.
\begin{defn}\rm
A map
\beq
 \pi \, \colon \, \mathbb{C}^n
\otimes \mathbb{C}^m \;\longrightarrow\; \mathbb{C}^n \otimes
\mathbb{C}^m
\eeq
is a quantum conditional probability (QCPO) iff
satisfies (1.6) and (1.7).
\end{defn}

For a given $\pi$ and any $\rho \in S (M_n)$ one can define a
compound state
\beq \omega \;=\; (\rho^{1/2} \otimes I_m)\, \pi\,
(\rho^{1/2} \otimes I_m)\,,
\eeq
which has the following properties
\beq
\mbox{tr}_{{\mathbb{C}^m}}\, \omega \;=\;\rho
\eeq
and
\beq
\mbox{tr}_{\mathbb{C}^n}\, \omega \;=\; \mbox{tr}_{\mathbb{C}^n}\, \pi
(\rho \otimes I_m) \;=\; \varphi^* (\rho)\,,
\eeq
where
\[
\varphi^* \, \colon \, S (M_n) \,\longrightarrow\, S (M_m)\, .
\]

The study of the relation between $\varphi^*$ and $\pi$ is based
on duality between quantum maps and composite states which has
been investigated in detail in [4,11] (see also references therein).

\section{Classification of Composite States and Positive Maps}
\setcounter{equation}{0}

It has been shown above that the construction of composite states is
based on the notion of quantum conditional probability operator
$(QCPO)\,$ $ \pi \,\colon \mathbb{C}^n \otimes \mathbb{C}^m \to
\mathbb{C}^n \otimes \mathbb{C}^m$. In what follows the case $m=n$
will be considered for simplicity.

Let ${\mathcal H}$ be a Hilbert space and ${\mathcal H}_1 =
{\mathcal H}_2 = {\mathcal H}$, the following order in the tensor
product ${\mathcal H} \otimes {\mathcal H} = {\mathcal H}_2
\otimes {\mathcal H}_1$ will be used, and the partial trace with
respect to the Hilbert space ${\mathcal H}_a$ will be denoted by
$\mbox{tr}_a$.

Let $\{ e_1, \ldots , e_n\}$ be a fixed orthonormal basis in
$\mathbb{C}^n$ and $e_{kl} = e_k (e_l, \,\cdot\,)$ be the
corresponding basis in $M_n$, then $a \in M_n$ can be written in the
form \beq a \;=\; \sum_{i,j=1}^n (e_i, a\,e_j) e_{ij} \;=\;
\sum_{i,j=1}^n\tr (a\, e_{ij}^*) e_{ij}\, , \eeq and the transpose
map $\mathcal{T} \,\colon  M_n \to M_n$ (with respect to the basis
$\{e_1, \ldots , e_n\})$ has the form \beq \mathcal{T} (a) \;=\;
\sum_{i,j=1}^n (e_j, ae_i) e_{ij} \;=\; \sum_{i,j =1}^n e_{ij}\, a\,
e_{ij} \, . \eeq A generic element $x \in \mathbb{C}^n \otimes
\mathbb{C}^n$ can be written in the form \beq x \;=\; \sum_{i=1}^n
x_i \otimes e_i \;=\; \sum_{i=1}^n (a\, e_i) \otimes e_i\,, \eeq
where $x_1, x_2, \ldots , x_n \in \mathbb{C}^n$ and $a \in M_n$.

With every $a \in M_n$, such that $\mbox{tr} \,(a^*\,a)=1$, one can
associate one-dimensional projections $p_a \,\colon
\mathbb{C}^n\otimes \mathbb{C}^n \to \mathbb{C}^n\otimes
\mathbb{C}^n$ \beq p_a \;=\; \sum_{i,j=1}^n a\, e_{ij}\, a^* \otimes
e_{ij} \,. \eeq Moreover, two projections $p_a$ and $p_b$ are
orthogonal provided that $\mbox{tr} \,(a^* b)=0$. As a consequence
of (2.4) any positive operator $\hat{A} \,\colon  \mathbb{C}^n
\otimes \mathbb{C}^n \to \mathbb{C}^n \otimes \mathbb{C}^n$ has the
form \beq \widehat{A} \;=\; \sum_{i,j=1}^n \sigma_{ij} \otimes
e_{ij} \;=\; \sum_{i,j=1}^n \varphi (e_{ij}) \otimes e_{ij}\,, \eeq
where \beq \varphi (e_{ij}) \;=\; \sum_{\alpha=1}^{n^2}
\lambda_\alpha\, a_\alpha \,e_{ij}\, a_\alpha^* \eeq and
\[
\lambda_\alpha \geq 0\, ,
\qquad\mbox{tr}\, (a_\alpha a_\beta^*) = \delta_{\alpha \beta}\, ,
 \]
i.e.~$\{a_1, a_2, \ldots , a_{n^2}\}$ is an
orthonormal basis in $M_n$.

Relation (2.5) can also be rewritten in the form \beq \hat{\sigma}
\;=\; (\varphi \otimes \mbox{id}) \sum_{i,j=1}^n e_{ij} \otimes
e_{ij} \,, \eeq which gives the relation between elements of $(M_n
\otimes M_n)^+$ and completely positive maps in $M_n$. In order to
classify the states on $M_n \otimes M_n$ it is convenient to
introduce the following cones in $M_n \otimes M_n$: \beq V_s \;=\;
\mbox{conv}\, \Big\{ \sum_{i,j=1}^n a\, e_{ij}\, a^* \otimes
e_{ij}\, \colon \; a \in M_n\,, \,\,\, \mbox{rank}\, a \leq s \Big\}
\eeq where conv$\,X$ means convex (not normalized) set generated by
elements of $X$, and \beq V^s \;=\; (\mbox{id} \otimes \mathcal{T})
V_s\,, \eeq where $\mathcal{T}$ is the transpose map on $M_n$, i.e.
\beq V^s \;=\; \mbox{conv}\, \Big\{ \sum_{i,j}^n a\, e_{ij}\, a^*
\otimes e_{ji}\, \colon\; a \in M_n\, , \,\,\,\mbox{rank}\, a \leq
s\Big\} \, . \eeq It follows from (2.8) and (2.9) that the following
chains of inclusions \beqn &  V_1 \subset V_2 \subset \ldots \subset
V_n , \qquad V^1 \subset V^2 \subset \ldots \subset
V^n\, ,& \nonumber \\
&  V_1 \cap V^1 \subset V_2 \cap V^2 \subset \ldots \subset V_n
\cap V^n & \eeqn
hold true.

It is clear that $V_n$ coincides with the cone $(M_n \otimes
M_n)^+$ of all positive semidefinite elements of $M_n \otimes
M_n$, $V_1 =V^1$ is the cone generated by elements $a \otimes b$,
where $a,b$ are positive elements of $M_n$, i.e.\ $V_1$ coincides
with separable (not normalized) states on $M_n \otimes M_n$, while
$V_n \cap V^n$ is the set of all (not normalized) PPT states on
$M_n \otimes M_n$ (by definition).

Using the results of  [4,\,5,\,6] the above cones can be used to
classify positive maps. Let $P_s$, $P^s$ and $P_s \cup P^r$ be the
cones of $s$-positive, $s$-copositive maps and sums of $s$-positive and
$r$-copositive ones, respectively. One can verify
that
\beqn
\varphi \in P_s &\Longleftrightarrow & (\varphi \otimes
\mbox{id}) V_s \in (M_n \otimes M_n)^+  \nonumber \\
\varphi \in P^s &\Longleftrightarrow & (\varphi \otimes \mbox{id})
V^s \in (M_n \otimes M_n)^+
\eeqn
and
\[
\varphi \in P_s \cup P^r \;\Longleftrightarrow\; (\varphi \otimes
\mbox{id}) V_s \cap V^r \in (M_n \otimes M_n)^+ \,.
\]

Relations (2.12) can be considered as an extension of the Horodecki
theorem [7] which gives the characterization of the cone $V_1$ in
terms of positive maps. It should be pointed out that our
knowledge concerning the above mentioned cones is rather poor. In
fact only the structure of cones $V_n$ and $V^n$ is  known.

In the $n$-dimensional case an example of an element $V_2 \cap
V^2$ which is not separable has been given in  [8].

The cones $V_r$ and $V^r$ and  $V_r \cap V^s$ can be used for classification of
completely positive maps.
\begin{defn}\rm
A completely positive map $\varphi\,\colon M_n\longrightarrow M_n$
is said to be $s$-{\it comple\-tely positive} if \beq
\sum^n_{i,j=1}\varphi(e_{ij})\otimes e_{ij}\;\in\; V_s \eeq and
$(r,s)$-{\it completely positive} if
\[
\sum^n_{i,j=1}\varphi(e_{ij})\otimes
e_{ij}\;\in\; V_r\otimes V^s\,.
\]
\end{defn}
Let us observe that the set $\mathcal{P}_{rs}$ of all $(r,s)$-completely
positive maps is a subset of $P_n\cap P^n$, the subset of maps which are
completely positive and completely copositive. On the other hand,
 $(r,s)$-completely positive maps generate PPT-states since the inclusion
 $V_r \cap V^s\subseteq V_n \cap V^n$ holds.

It is also convenient to consider $s$-completely positive maps which
are $k$-coposi\-tive, i.e.~elements of the set $P_n\cap P^k$
($k<n$) which generate NPT states.
The construction of composite states in terms of QCPO will
be used to find out some classes of PPT and NPT states.


Indeed, taking into account (2.5), one finds out that the general
form of $\pi$ is given by
\beq
\pi \;=\; \sum_{i,j}^n \varphi (e_{ij}
)\otimes e_{ij}\,,
\eeq
where $\varphi$ is a completely positive
normalized map in $M_n$, i.e., $\varphi(I_n)=I_n$. It follows from (2.14) and (1.8) that the
composite state can be written in the form
\beq
\omega \;=\;
\sum_{i,j=1}^n \rho^{1/2} \, \varphi (e_{ij})\,\rho^{1/2} \, \otimes e_{ij}
\eeq
and
the relations
\beqn
\mbox{tr}_1 \, \omega &=& \rho\,, \nonumber \\
\mbox{tr}_2 \, \omega &=& (\mathcal{T} \circ \varphi^*)(\rho) \;=\;
\psi^* (\rho)\, , \eeqn
hold true.

The dual map
$\psi$ can be written as
\beq
\psi(e_{ij}) \;=\; \varphi (e_{ji}) \;=\;
(\varphi\circ {\mathcal{T}})(e_{ij})
 \quad {\rm or}\quad
\psi^\ast\;=\;{\mathcal{T}}\circ\varphi^\ast \,.
\eeq
The properties of the composite state  $\omega$ can be summarized as follows.
\begin{cor}\rm
The composite state  $\omega$ is a PPT iff $\varphi$ is $(r,s)$-completely positive.
\end{cor}
\begin{cor}\rm
The composite state  $\omega$ is NPT iff $\varphi$ is $r$-completely positive
and $k$-copositive $(k<n)$ provided
rank$\,\rho=n$.
\end{cor}

The above  results imply that the
construction of PPT-states is reduced to normalized completely
positive and completely copositive maps, while, entangled but not
PPT-states are induced by normalized completely positive and
$k$-completely copositive maps.

Next we will analyze some examples of normalized completely positive
and completely copositive maps.
\begin{exmp}\rm
Let us consider the following QCPO:
\beq
\pi_\lambda \;=\; \frac{1-
\lambda }{n} \, I_n \otimes I_n + \lambda \sum_{i,j=1}^n e_{ij}
\otimes e_{ij}\,,
\eeq
where
\beq
- \frac{1}{n^2 - 1} \;\leq\; \lambda
\;\leq\; \frac{1}{n+1} \, .
\eeq
The above $\pi$ is, up to
normalization, the Horodecki state [8], and for $\lambda$
satisfying (2.19) $\pi$ is separable, i.e., $\pi_\lambda\in V^1\cap V_1$.

Let $\psi \,\colon  M_n \to M_n$ be a normalized positive map, then
\beq \sigma_\lambda \;=\; (\psi \otimes \mbox{id})\, \pi_\lambda
\eeq is a QCPO which is separable, and the relation \beq \sum_{ij}
\varphi_\lambda(e_{ij}) \otimes e_{ij} \;=\; (\psi \otimes
\mbox{id}) \, \pi_\lambda \eeq defines completely positive and
completely copositive maps of the form \beq \varphi_\lambda (e_{ij})
\;=\; \frac{1 - \lambda}{n}\, I_n\, \delta_{ij} + \lambda \,\psi
(e_{ij})\,, \eeq which is $(1,1)$-completely positive.
\end{exmp}
\begin{exmp}\rm
Let us consider the following QPCO: \beq \pi_\gamma \;=\;
N_\gamma^{-1} \Bigl( n\, I_n \otimes I_n + \sum_{i,j=1}^n a_{ij}
\otimes e_{ij} \Bigr)\,, \eeq where \beqn
a_{ij} &=& n\, e_{ij}\,,\qquad i \neq j\,, \\
a_{ij} &=& \Bigl( 1 - \frac{1}{\gamma^2} \Bigr) (\gamma^2
e_{i+1, i+1} - e_{n+i-1, n+i-1} )\,\, (\mbox{mod}\,n)\,, \\
N_\gamma &=& n^2 + \Bigl( 1  - \frac{1}{\gamma^2} \Bigr) (\gamma^2 -
1)\,, \qquad \gamma^2 >0\, . \eeqn It has been shown in  [4] that
$\pi_\gamma \in V_2 \cap V^2$ but it is not separable.

The relation
\beq
\sum_{i,j =1}^n \varphi_\gamma (e_{ij}) \otimes e_{ij}
\;=\; \pi_\gamma
\eeq
defines a normalized  $(2,2)$-completely positive
 map $\varphi$ which has the form
\beq
\varphi_\gamma (e_{ij}) \;=\;  \frac{1}{N_\gamma} (n\, I_n\, \delta_{ij} +
a_{ij})\, .
\eeq
\end{exmp}
\begin{exmp}\rm
The map $\varphi$ given by (2.28) has the following form
\beq
\varphi (a) \;=\; \sum_{i,j}^n c_{ij} \, e_{ij}\, a\,e_{ij}^* + \mu a \,,
\eeq
where
\beq
\varphi(I_n)\;=\;\sum_{i=1}^n e_{ii}\Big(\sum^n_{j=1}c_{ij}+\mu\Big)\,.
\eeq
Taking into account
(2.2) one finds that the relation
\beq
\varphi (\mathcal{T} (a)) \;=\;
\mathcal{T}(\varphi (a))
\eeq
holds.
\end{exmp}
\begin{thm}
The map $\varphi$, as in (2.29), is completely positive iff the
following conditions \beq c_{ij} \;\geq\; 0\,,\qquad i \neq j \eeq
and \beq \bigl[ c_{ii} \delta_{ij} + \mu \bigr] \;\geq\; 0 \eeq are
satisfied.
\end{thm}
{\it Proof}.\quad
The map $\varphi$ can be written in the form \beq \varphi (a) \;=\;
\sum_{i,j=1}^n (\delta_{ij} c_{ii} + \mu) e_{ii}\, a\, e_{jj} +
\sum_{i\neq j} c_{jj}\, e_{ij}\, a\, e_{ij}^* \eeq on the other hand
the map $\psi$ is completely positive iff it has the form \beq
\psi(a) \;=\; \sum_{\alpha, \beta =1}^{n^2} \lambda_{\alpha \beta}
f_\alpha a f_\beta^*\,, \eeq where \beq \mbox{tr} \, (f_\alpha
f_\beta^*) \;=\; \delta_{\alpha \beta} \eeq and \beq
\bigl[\lambda_{\alpha \beta} \bigr] \;\geq\; 0\, . \eeq Taking into
account (2.35)--(2.37) and (2.34), one finds conditions (2.33) and
(2.34).
\begin{thm}
The map $\varphi$ is completely copositive iff the following
conditions
\beqn
c_{ii} + \mu &\geq& 0\,, \\
c_{ij} + c_{ji} &\geq& 2 \mu\,,\qquad i \neq j\,, \\
c_{ij} + c_{ji} &\geq& - 2 \mu\,,\,\quad i \neq j\,, \\
c_{ij} c_{ji} &\geq& \mu^2\,,\qquad i \neq j
\eeqn
are
satisfied.
\end{thm}
{\it Proof.}\quad
The map $\varphi$ is completely copositive iff the map $\mathcal{T}
\varphi$ is completely positive.

Taking into account (2.2) and (2.29) one finds \beq
(\mathcal{T}\circ \varphi) (a) \;=\; \sum_{i=1}^n (c_{ii} + \mu)
e_{ii}\, a\, e_{ii} + \sum_{i \neq j} (c_{ij}\, e_{ij}\, a \,
e_{ij}^* + \mu e_{ij}\, a\,  e_{ij} )\,. \eeq From (2.42) one finds
the condition (2.38). Let us introduce trace orthonormal operators
\beqn f_{ij} &=&
\frac{1}{\sqrt{2}} (e_{ij} + e_{ji})\,, \nonumber \\
g_{ij} &=& \frac{-i}{\sqrt{2}} (e_{ij} - e_{ji})\,,\qquad i <j\,.
\eeqn Then the equality \beqn & & \hspace*{-20mm} \sum_{i \neq j}
(c_{ij}\, e_{ij}\, a\, e_{ij}^* + \mu e_{ij}\, a\, e_{ij}) \;=\;
\sum_{k < l} \Big\{ \Big( \frac{1}{2} (c_{kl} + c_{lk}) + \mu \Big)
f_{kl}\, a\, f_{kl}
\nonumber \\
& & \qquad\qquad +\; \Big( \frac{1}{2}(c_{kl} + c_{lk}) - \mu \Big)
g_{kl}\, a\, g_{kl} - \frac{i}{2} (c_{kl} - c_{lk}) f_{kl}\, a\,
g_{kl}
\nonumber \\
& & \qquad\qquad\qquad\qquad +\;  \frac{i}{2} (c_{kl} - c_{lk}) g_{kl}\, a\, f_{kl}
\Big\}
\eeqn
holds true.

Taking into account (2.35) and (2.37) one finds that $\mathcal{T}
\circ \varphi$ is completely positive iff the matrices \beq
\Delta_{kl} \;=\; \left[ \begin{array}{ll} {\dST \frac{1}{2} (C_{kl}
+ C_{lk}) + \mu}~~~~~~~ & {\dST - \frac{i}{2} (C_{kl} - C_{lk})} \\
 & \\
{\dST \frac{i}{2} (C_{kl} - C_{lk})}~~~~~~~ & {\dST  \frac{1}{2}
(C_{kl} + C_{lk}) - \mu} \end{array} \right] \eeq are semipositive
definite. The conditions (2.39)--(2.41) are equivalent to
$\Delta_{kl} \geq 0$.
\begin{cor}\rm
The map
\beq
\varphi(a)\;=\;\sum_{i,j=1}^n c_{ij}\, e_{ij}\, a\,
e_{ij}^* + \mu  a
\eeq
is completely positive and completely copositive, i.e. $\varphi\in P_n\cap P^n$ iff the
following conditions
\beqn
c_{ij} &>& 0\,,\\
c_{ij} c_{ji} &\geq& \mu^2\,,\qquad i \neq j\,, \\
c_{ii} + \mu &\geq& 0\,, \\
\bigl[c_{ii}\delta_{ij} + \mu\bigr] &\geq& 0 \eeqn are satisfied.

The $\varphi$ is not normalized, but since $\varphi(I_n)>0$, the map
\beq
\psi(a)\;=\;\varphi(I_n)^{-1/2}\, \varphi(a)\,\varphi(I_n)^{-1/2}
\eeq
is normalized, and the state
\beq
\sum_{i,j=1}^n \rho^{1/2} \psi(e_{ij})\rho^{1/2} \otimes e_{ij}
\eeq
is a PPT state.
\end{cor}
\begin{cor}\rm
It follows from Theorems 1 and 2 that choosing $c_{ij}=k$, $1\leq k<
n$, and $\mu=-1$ the map \beqn \varphi_k(a) &=& (k-1)\Big(
\sum_{i=1}^n  e_{ii}\, a\, e_{ii} + \sum_{i<j}  f_{ij}\, a\,
f_{ij}\Big) + (k+1) \sum_{i<j}  g_{ij}\, a\, g_{ij} \eeqn is
completely positive, while the map \beq (\tau \circ
\varphi_k)(a)\;=\;k I_n \, ({\rm tr}\, a) -a \eeq is known to be
$k$-positive but not $(k+1)$-positive; i.e., $\varphi_k$ is
 completely positive and $k$-copositive, and consequently the state
\beq
\frac{1}{kn-1}
\sum_{i,j=1}^n \rho^{1/2} \, \varphi_k (e_{ij})\,\rho^{1/2} \, \otimes e_{ij}
\eeq
is NPT (by definition) provided rank$\rho=n$.
\end{cor}

Using Theorems 1 and 2 specialized for the case $c_{ii}=c$ and $\mu=-1$.
\begin{cor}\rm
The map
\beqn \varphi(a) &= & \sum_{i=1}^n (c-1) e_{ii}\, a\, e_{ii}+
\sum_{i < j} \Big\{ \Big( \frac{1}{2} (c_{ij} + c_{ji}) -1 \Big)
f_{ij} a f_{ij}
 \\
&+ & \Big( \frac{1}{2}(c_{ij} + c_{ji}) +1 \Big) g_{ij}\, a\, g_{ij}
- \frac{i}{2} (c_{ij} - c_{ji}) f_{ij}\, a\, g_{ij}
 +\frac{i}{2} (c_{ij} - c_{ji}) g_{ij}\, a\, f_{ij}
\Big\} \nonumber
\eeqn
is completely positive iff $c\geq 1$ and $c_{ij} c_{ji}\geq 1\quad i\neq j$.

On the other hand
\beqn
(\tau \circ \varphi)(a)\;=\;
c \sum_{i=1}^{n-1} h_i\, a\, h_i + \sum_{i\neq j}c_{ij}\, e_{ij}\, a\, e_{ij}^\ast -\frac{n-c}{n}a\,,
\eeqn
where
$h_1,h_2,\ldots h_{n-1}\in M_n$, $h_i=h_i^\ast$, ${\rm tr}\, h_i=0$,
${\rm tr}\, h_i h_j=\delta_{ij}$
and $ {\rm tr}\, h_i f_{kl}={\rm tr}\, h_i g_{kl}=0$.
Using the result of [9] one finds that $(\tau \circ \varphi )$ is $k$-positive but not
$(k+1)$-positive provided the following conditions are satisfied \cite{tt}
\beqn
& 1\;\leq\; k \;\leq\; c \;<\; k+1 \;<\; n\,, & \\
&\dST\frac{k}{n-k} \;\leq\; \frac{c_{ij}}{n-c}\;<\;\frac{k+1}{n-k-1}\,.&
\eeqn
In this case the map $\varphi$ generates NPT states.
\end{cor}

\section*{Acknowledgments}
The work of M.\,A. and G.\,M. has been partially supported by a cooperation grant INFN-CICYT.
The work of M.\,A. has also been partially supported by the Spanish MCyT grant
FPA2000-1252.  A.\,K. has been supported by the Grant PBZ-MIN-008/P03/2003.


\end{document}